# IDENTIFYING AND ANALYSIS OF SCENE MINING METHODS BEASED ON SCENES EXTRACTED FEATURES

ASHRAF SADAT JABARI

Department of Computer Engineering, Islamic Azad University,
Qazvin Branch, Qazvin, Iran
a.jabari@Qiau.ac.ir

MOHAMMADREZA KEYVANPOUR

Department of Computer Engineering, Alzahra University,
Tehran, Iran
Keyvanpour@Alzahra.ac.ir

Abstract :
Scene mining is a subset of image mining in which scenes are classified to a distinct set of classes based on analysis of their content. In other word in scene mining, a label is given to visual content of scene, for example, mountain, beach. Scene mining is used in applications such as medicine, movie, information retrieval, computer vision, recognition of traffic scene. Reviewing of represented methods shows there are various methods in scene mining. Scene mining applications extension and existence of various scenes, make comparison of methods hard. Scene mining can be followed by identifying scene mining components and representing a framework to analyzing and evaluating methods. In this paper, at first, components of scene mining are introduced, then a framework based on extracted features of scene is represented to classify scene mining methods. Finally, these methods are analyzed and evaluated via a proposal framework.

## 1.  Introduction

Scene mining is grouping scenes based on analyzing of contents [3]. In scene mining it is determined which classes a scene is belong to.

Scene mining is used in applications such as medicine, to better medicos recognition[8], Traffic scenes, to recognizing dynamic objects of scene[5], Information retrieval system, to decreasing semantic gap and improving  organization of databases[9,11].

Analysis and evaluation of scene mining methods are hard because of applications and various methods of scene mining, therefore it is essential to representing a general classification for scene mining methods. This paper represents a framework to classifying scene mining methods. The rest of the paper is organized as follows. In session 2 we review related work in scene mining and then identify scene mining. Session 3 represents general structure of scene mining process and introduces primary components of scene mining. Session 4 describes a proposed framework according to various methods of scene mining and session 5 analyzes and evaluates classified methods of scene mining according to proposed framework.

## 2.  Related work

Semantic scene classification is categorizing scenes into a discrete set of classes. In scene mining is determined which semantic class, a scene is belong to (sky, street, beach). Scene mining groups scenes into discrete semantic set of classes based on their content analysis. Vogel in [12] tells that scene mining is sorting scenes into a set of semantic groups such beach, sky. In [3] is told, in scene mining a label is given to visual content of special scene. Scene mining places a scene automatically into one of a set of physical (e.g., indoor/outdoor, orientation) or semantic categories (e.g., beach or party). In general, scene mining is automatically categorizing scenes into a discrete set of semantic classes.

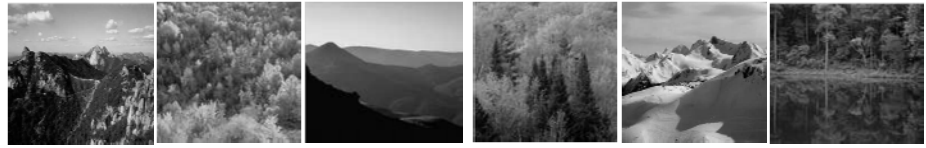

Figure1.a. Forest and mountain scenes before scene mining

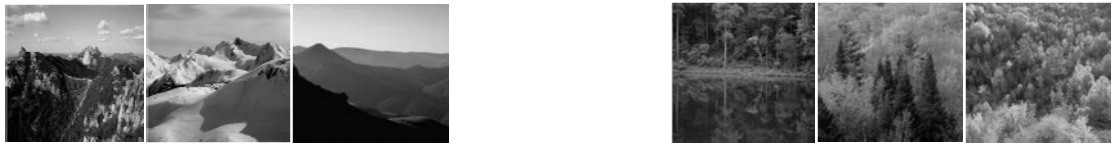

Mountain class                                                            Forest class

Figure 1.b. Mountain and forest scenes after scene mining

In Figure 1.a have been shown some forest and mountain scenes before scene mining. Figure 1.b shows forest and mountain scenes after scene mining. In scene mining, at first, a vector is created based on scenes extracted features, then the scene is represented and finally a classifier determines which classes the scene is belong to. [3] tells object recognition is essential in scene mining because objects should be recognized in scene understanding, also scenes can be classified without full knowledge of every object in the scene. For instance, if a person recognizes trees at the top of a scene, grass on the bottom, and people in the middle, he may hypothesize that he is looking at a park scene, even if he con not see every details in the scene.

Scene mining can be used in image and video retrieval and it is effective in browsing databases and decreasing semantic gap in image content. Increasing precision and reducing time of search are result of scene mining in this application [9, 11]. [14] has used audio classification in video scene mining because audio computation is lower and it is possible to doing more experiments based on visual information. In medicine, scene mining helps medicos to compare their recognition to Endoscopic scenes in databases and increases their recognition precision [8]. Scene mining in traffic scenes causes recognizing both the road type (straight, left/right curve) as well as a set of encountered objects(car, pedestrian)[5]. Scene mining applications in various fields can be effective in identifying and analyzing scene mining components. In next session, scene mining process and its components are introduced.

## 3. Scene mining process

Scene mining is grouping scenes to a discrete set of classes. In this session, we define formal form of scene mining process and then show components of it. We define a fundamental assumption for scene mining as follow:

$$\{ \forall\ s_i \in S\ \exists\ c_T \in C\ ,\ s_i\ \Delta\ c_T \} \qquad (1)$$

Where C is a set of classes and S is a set of scenes. $s_i$ and $c_T$ are input scene and target class respectively and $\Delta$ shows that $s_i$ belongs to $c_T$ and $i, T \in \{1, 2, 3, …, n\}$. This assumption shows, for every scene there is a class that the scene belongs to it. Now, we show formal form of scene mining process according to above assumption :

$$c_T = \underset{s_i \in S,\ c_T \in C}{\operatorname{argmax}}\ F(s_i, c_j) \qquad (2)$$

where $c_j$ , $c_T \in C$ and $s_i \in S$ and $F(s_i, c_j)$ computes similarity of $s_i$ and $c_j$. It means there is a function such as F which assigns $c_j$ to $s_i$ and then shows value of similarity. $c_T$ is a target class that $s_i$ has maximum similarity to members of it.

According to definition of scene mining process, it consists of five primary components: scene database, preprocessing, feature extraction, scene mining, target class. Figure 2 shows stages of scene mining process. In continuance,  a detailed description of all components are provided.

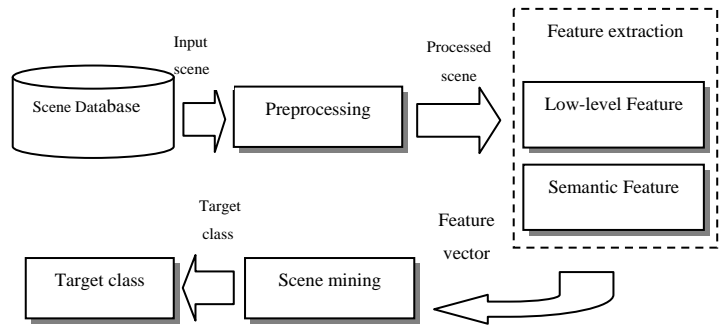

Figure2. General Process of Scene Mining

### 3.1. Scene database

The first component of scene mining process is scene database. Scene database consists of scenes samples to training and testing classifiers. These two groups of samples are used based on used algorithms in scene mining. Testing samples are entered in preprocessing stage to preprocessing process.

### 3.2. Preprocessing

The second stage of scene mining process is preprocessing. This stage consists of sub stages such as normalization, quantization, sampling, equalization. These processes are done to improve input scene quality. After this stage, processed scene is entered in feature extraction stage to creating feature vector.

### 3.3. Feature extraction

The third stage of scene mining process is scene features extraction. In this stage, two groups of features are extracted based on scene mining algorithms. 1- low-level features; Color, texture, edge are low-level features. There are many models to extract color such as HIS, RGB, CYM. To extract texture, Wavelet transform, filtering methods can be used. 2- Semantic features; Scene mining based on low-level features con not be useful in all applications and does not have enough precision, therefore scene mining precision and its domain of applications can be increased by semantic feature extraction.

### 3.4. Scene Mining

The forth component of scene mining, is mining of a scene. It is based on machine learning algorithms to determine classes of input scenes. Classification, clustering, association rules are methods that are used to mining a scene. SVM (Support Vector Machine) is used in some scene mining processes. This is defined as follow:

$$F(x) = \sum_i \alpha_i y_i K(x, x_i) + b \qquad (3)$$

$\alpha_i$ is function parameter, X is unlabeled sample, $X_i$ is support vector and $K(x, x_i)$ is kernel function. This stage may use combination of algorithms. [14] at first, has classified audio by classification method and then has mined tennis scenes by clustering algorithms. Final it has used synchronization of audio and scene to mining tennis scenes.

### 3.5. Target class

Final stage of scene mining is determining class of input scene. In this stage, the input scene is classified into a target class after it is passing preprocessing, feature extraction stages and after it is mined via a machine learning algorithm.

## 4. Proposed framework to analyze and evaluate extracted-features based scene mining methods

In a scene mining system, at first, a vector is created using extracted features from various sources. After providing a set of features, a classifier determines which class the input scene is belong to.

By reviewing existence methods in scene mining and extracted features, scene mining methods can be categorized into 4 groups based on extracted features: scene mining based on low-level features, scene mining based on low- level semantic features, scene mining based on middle- level semantic features, scene mining based on high- level semantic features. In figure 3, this classification is shown.

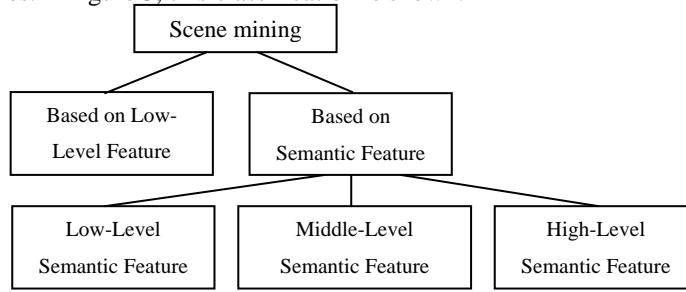

Figure3.Classification of scene mining methods based on scenes extracted features.

### 4.1. Scene mining based on low-level features

Color, texture and edge are used in scene mining as low-level features. In this method, at first a vector is created that consists of low-level features, then a classifier classifies the scenes. The models such as RGB, LAB, HIS, $YC_rC_b$ are used in color feature extraction. Each model can be transformed to others. For instance, following equations can be used to transform RGB model to HIS model:

$$H= \begin{cases} \theta & \text{if} \quad B <= G \\ 360\text{-} \theta & \text{if} \quad B > G \end{cases} \qquad (4)$$

$$S = 1 - (\ 3 / (\ R + G + B\ )\ [\min(\ R\ ,\ G\ ,\ B\ )\ ]\ ) \qquad (5)$$

$$I = 1 / 3\ (R + G + B) \qquad (6)$$

Texture shows scene content and can be extracted via Gabor Gray-level Co-Occurrence Matrix, Wavelet transform and Filtering. Co-Occurrence Matrix is a matrix or distribution that is defined over an image and shows distribution of occurrence values. Co-Occurrence matrix over an M*N image, parameterized by ($\Delta$ x, $\Delta$ y) offset is defined as follows:

$$C_{\Delta x,\Delta y}(i,j)=\sum_{p=1}^{n}\sum_{q=1}^{m}\cdot \begin{cases} 1 & \text{, if } I(p,q) = i \text{ and } I(p + \Delta x, q + \Delta y) = j \\ 0 & \text{,} \qquad \text{otherwise} \end{cases} \qquad (7)$$

The value of the image originally referred to the grayscale value of the specified pixel. Scene mining based on low-level features con not be used in all applications. For instance, querying a scene by color only sometimes gives understandable, but unmeaningful results. In figure 4 is shown a query for a Ferrari that can return a rose, if two images are most similar in global color distribution [3].

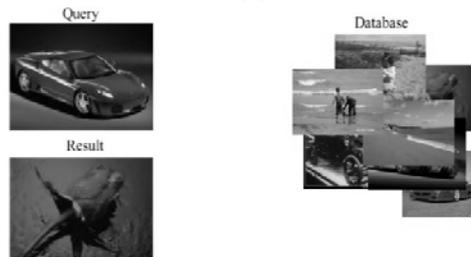

Figure4.Unmeaningful result of Ferrari query based on color feature [3].

## 4.2. Scene mining based on semantic features

The second method in scene mining based on scene extracted features is scene mining based on semantic features. In this method, not only low-level features are used but also semantic features are used to creating feature vector, such as person motions type, person standing direction, counts of existence person in scene. In next session, scene mining methods based on semantic features are introduced.

### 4.2.1. Scene mining based on low-level semantic features

In this method, combination of low-level features and low-level semantic features are used. Object recognition [2], detectors and descriptors [6] and pattern recognition are used to extracting semantic features. Because of unrelated objects and scenes ambiguity, this method is not useful in all applications. Existence of unrelated objects causes to decrease precision in scene mining and belonging of scenes objects to multi classes makes scenes ambiguous [7].

### 4.2.2. Scene mining based on middle – level semantic features

In some applications, scene mining is based on combination of low-level features and semantic features which are extracted from other sources such as text and audio. [4] has used video and text in sport scene mining that its feature vector consists of textual and low-level features. In the game action, a complete description of playing game is instantly produced in sentences and archived for later use. Each sentences can be represented as: <sentence time, sentence content>. For each replay scene, those sentences within the window $<T_{replay} - T_l$ , $T_{replay}-T_r>$ are selected for textural feature extraction. $T_{replay}$ denotes the ending time of a replay and $T_l$ and $T_r$ denote the left boundary and right boundary of the associated window respectively. Table 1 shows semantic event terms and their synonymous terms of soccer sport.

Table1. Semantic event terms and synonymous[4]

| Replay Category | Correlative terms |
|---|---|
| GR | G-O-A-L, goal |
| SR | shoot, smash, blast, chip, fire, crack, ranger, head, shot |
| AR | corner, punch, collect, save, clear, cross, goalkeeper, box, goal, |
| FR | free kick, yellow card, book, red card, medical, injured, treatment, stretcher, go off, return, foul |
| OR | off side |
| OBR | throw in, goal kick, corner |
| OTR | N.A. |

The term-frequency vector $X_i$ of a replay is defined as:

$$X_i=[X_{1i},X_{2i},....,X_{3i}, X_{Mi}]^T \qquad (8)$$

Where $X_{ji}$ denotes the frequency of term f $_i$ $\in$W in the related sentences of the i th replay W={f$_1$, f$_2$ , … , f$_n$} where W denotes the complete related vocabulary set of selected semantic event terms and their synonymous terms. Finally a vector which consists of visual features and textual features, is created. In [14] two vectors are created , first one consists of audio features and second one is built by video features. Then both of them are mined separately and finally, it uses synchronization of audio and image in video to mining racquet sports scenes.
 Limitation of resources to extracting features is this method challenge. Nonexistence of discrete audio in tennis video is instance of this challenge [14].

### 4.2.3. Scene mining based on high-level semantic features

Scene mining based on high-level semantic features uses semantic features based on its application. For example [1] uses 3-dimentional model to extracting features. In figure 5 has been shown 3-dimentional scenes which have been used in crime scene mining. It uses features such as type of used weapon, distance of the weapon in relation to the victim, angles between the person and the extremities in crime scene mining.

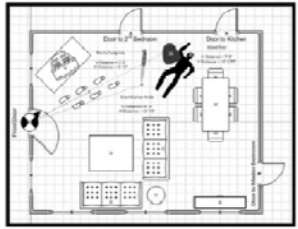

Figure 5. 3-D Model of crime scene[1]

In [13] is used motion features in movie scene mining. Each frame is divided into blocks with the size of M*N pixels, and each block has a motion vector. The motion vector is estimated by searching for the best matching block within a search window of size P in next frame, P is shown in below equation:

$$P = (M+2Wx)*(N+2Wy) \qquad (9)$$

In figure 6, counts of persons in scene, direction person face and state of persons face are shown as semantic features of movie scenes.

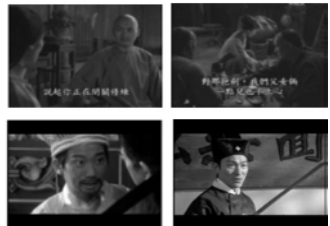

Figure6. Semantic features in movie scenes[13].

## 5. Proposed framework to analyzing scene mining methods

In this session, 4 metrics are proposed to analyzing scene mining methods, such as: identity of feature resource, counts of feature vectors, complexity, reducing scene ambiguity. Next session analyzes and evaluates scene mining methods based on proposed metrics.

### 5.1. Identity of resources to extracting features

Identity of resources is the first proposed metric in extracting features and creating feature vectors to analyzing scene mining methods. In scene mining based on low-level features, only images are used, so the identity of resources is image in this method. In scene mining based on low-level semantic features, features can be extracted from image and video. In scene mining based on middle-level semantic features, other resources such as text and audio are added to image and video. In the last method, resources are depend on applications and features are extracted based on special application.

### 5.2. Count of feature vectors

Count of feature vectors is the second metric in analyzing of scene mining methods. In the first and second methods of scene mining, a vector is created by extracted features to mining the scene. In the third and the forth methods, it is possible that discrete feature vectors are created based on extracted features of per resource and each vector is classified discretely. [14] has extracted a audio features vector and a video features vector and then has classified them discretely to sport scene mining. It is possible, a vector is created based on extracted features from various resources and then scenes are mined [4].

### 5.3. Complexity

The other proposed metric to analyzing scene mining methods is complexity. How features extract to creating vector, count and types of classifiers in scene mining are metrics which affect complexity of scene mining. With regard to the fact that in four scene mining, methods can use multi classifiers and the first method creates one vector, the first method has the lowest complexity. Also in scene mining based on low-level semantic features is created one vector, but the components of its vector have semantic identity. So its complexity is more than the first method. In Scene mining based on middle and high–level semantic features, not only there are some vectors but also identity of vectors components are various. So these methods have more complexity in comparison with two first methods.

### 5.4. Reducing scene ambiguity

The forth metric to analyzing scene mining methods is reducing scene ambiguity. Some scenes have objects which are related to multi classes. It causes, a scene classifies to multi classes and accuracy is decreased. In the first method because of using only low-level features and in the third method because of using resources which are not image and video, there is not enough ability to recognizing scenes objects and reducing semantic gap. In scene mining based on low-level semantic features because of using methods such as object recognition and pattern recognition, scene mining works better than the first and second methods. In forth method, high-level semantic features cause better performance to mining scenes. In table 2 metrics of scene mining evaluation based on extracted features are shown.

Table2. Evaluation of scene mining methods based on extracted features

| Applications | Reducing scene ambiguity | Complexity | Count of feature vector | Identify of feature resource | Metrics | Methods |
|---|---|---|---|---|---|---|
| Land cover images[10], Medicine[8] | Weak | Low | 1 | Image | Scene mining based on low-level features | |
| Information Retrieval[11], Traffic scenes[5] | Middle | Middle | 1 | Image, Video | Low-level features | Scene mining based on semantic features |
| Sport[4,14] | Partly weak | High | N | Image, Video, Text, Audio | Middle-level features | |
| Crime Scene Recognition[1], Movie[13] | Partly acceptable | High | N | Depends on application | High-level features | |

### References

[1]   Abu hana, R.O.; Freitas, C.O; Oliveira, L.S; Bortolozzi, F.(2008): Crime scene representation (2D, 2D, stereoscopic projection) and classification, Journal of universal computer science, 14(18), pp.2953-2966.
[2]   Bosch, A.; Mu˜noz, X.; Oliver, A.; Mart´, R.(2006): Object and Scene Classification: what does a Supervised Approach Provide us?, Proceedings of the 18th International Conference on Pattern Recognition (ICPR'06).
[3]   Boutell, M.(2005): Exploiting context for semantic scene Classification, Technical Report894, University of Rochester, New York.
[4]   Dai, J.; duan, l.; Tong, X.; Xu, Ch.; Tian, Q.; Lu, H.; Jin, J.S.(2005): Replay scene classification in soccer video using Web Broad cast text, IEEE 0-7803-9332-5.
[5]   Es ; Muller, A. ; Grabner, T.; Gool , H.(2009): Segmentation-based Urban Traffic scene understanding, ESSET Al.
[6]   Gokalp, D.; Aksoy, S. (2005): Scene classification using Bag-of-region representation, Bilkent 06800 Ankara ,turkey.
[7]   Grossberg, S.; Huang, T.(2009): Artscene: A Neural system for natural scene classification, Journal of vision, 9(4), pp.1-19.
[8]   Guillou, C.Le.; Cauvin, J- M.; Solaiman, B.; Robaszkiewicz, M. (2001): Upper Digestive Endoscopic Scene Analyze, Proceedings of the 23rd Annual EMBS International Conference.
[9]   Kimond, W.; Park, J.; Kim, Ch.(2009): A novel method for efficient indoor-outdoor image classification, J sign process syst, DOI 10.1007/S11265-009-0446-0.
[10]  Knorn, J.; Rabe, A.; Radeloff, V.C.; Kuemmerle, T.; Kozak, J.; Hostert, P.(2009): Land cover mapping on large area using chain classification of neighboring landsat satellite images, Ramote sensing of Enviroment ,Rse-0733.
[11]  Luo, J.; Boutell, M.; Gray, R.T.; Brown, Ch. (2005): Image transform bootstrapping and its application to semantic scene classification, IEEE Transactions on system, Man and Cybernetics-Partb, 35(3) , pp. 563-570.
[12]  Vogel, J.(2003): Annotated bibliography content–based image retrieval: performance evaluation and semantic scene understanding.
[13]  Wang, Y.; Chang, Ch.(2003): Movie scene classification using Hidden Markov model, 16th IPPR conference on computer vision, Graphics and image processing.
[14]  Xing, L.; ye, Q.; Zhang, W.; Huang, Q.; Yu, H.(2005): A scheme for racquet sports video analysis with the combination of audio-visual information , 59(60), pp.259-266.